*C*-axis electrical resistivity of PrO$_{1-a}$F$_a$BiS$_2$ single crystals

Masanori Nagao[1,2*], Akira Miura[3], Satoshi Watauchi[1], Yoshihiko Takano[2], and Isao Tanaka[1]

[1]University of Yamanashi, Kofu 400-8511, Japan

[2]National Institute for Materials Science, Tsukuba 305-0047, Japan

[3]Hokkaido University, Sapporo 060-8628, Japan

*E-mail : mnagao@yamanashi.ac.jp

**Abstract**

The high anisotropy in RO$_{1-a}$F$_a$BiS$_2$ (R denotes a rare-earth element) superconductors demonstrates their potential use as intrinsic Josephson junctions, considering the weak coupling among BiS$_2$-PrO(F)-BiS$_2$ (superconducting-normal-superconducting) layers along the *c*-axis. We grew PrO$_{1-a}$F$_a$BiS$_2$ single crystals using CsCl/KCl flux. The superconducting anisotropies of the grown single crystals were estimated to be approximately 40–50 from the effective mass model. The *c*-axis transport properties were characterized using single-crystal s-shaped intrinsic Josephson junctions with a focused ion beam. Along the *c*-axis, the crystals showed zero resistivity at 2.7 K and a critical current density of $1.33\times10^3$ A/cm$^2$ at 2.0 K. The current–voltage curve along the *c*-axis displayed hysteresis. The *c*-axis transport measurements under a magnetic field parallel to the *ab*-plane revealed a "lock-in" state due to the Josephson vortex flow, indicating that BiS$_2$ superconductors are promising candidates for intrinsic Josephson junctions.

**Main Text**

**1. Introduction**

The newly developed BiS$_2$-based layered superconductors Bi$_4$O$_4$S$_3$[1,2] and RO$_{1-a}$F$_a$BiS$_2$ (R = La, Ce, Pr, Nd, or Yb)[3-12] have attracted much interest. Superconductivity is induced by carrier doping through the substitution of F at the O site in ROBiS$_2$. Recently, we have reported on the growth of RO$_{1-a}$F$_a$BiS$_2$ (R = La, Ce, Nd) single crystals by using the CsCl/KCl flux method. Their structures and transport properties along the *ab*-plane were examined.[13-17] The superconducting anisotropies of RO$_{1-a}$F$_a$BiS$_2$ (R = La, Nd) single crystals have been estimated to be as high as 30 using an effective mass model,[13-15] which demonstrates their potential use as intrinsic Josephson junctions along the *c*-axis.[18,19] Intrinsic Josephson junctions were discovered in high-$T_c$ cuprate superconductors. Such junctions emerged in the crystal structures of layered superconductors. Consequently, devices based on self-assembled nano junctions have been achieved. The applications of intrinsic Josephson junctions include terahertz radiation sources[20] and quantum bits.[21,22] However, there have been no previous experimental reports on intrinsic Josephson junctions along the *c*-axis of BiS$_2$-based layered superconductors. Thus, their *c*-axis transport properties have attracted much attention. Additionally, single crystals of $RO_{1-a}F_aBiS_2$ (R = Pr) have not yet been studied.

In this investigation, we grew $PrO_{1-a}F_aBiS_2$ single crystals with different F concentrations using CsCl/KCl flux and examined their transport properties parallel and perpendicular to the $c$-axis. Furthermore, we evaluated the s-shaped junctions[23,24] along the $c$-axis of a $PrO_{1-a}F_aBiS_2$ single crystal.

## 2. Experimental procedure

Single crystals of $PrO_{1-a}F_aBiS_2$ were grown using CsCl/KCl flux.[13,15] The raw materials of $Pr_2S_3$, Bi, $Bi_2S_3$, $Bi_2O_3$, and $BiF_3$ were weighed with a nominal composition of $PrO_{1-x}F_xBiS_2$ ($x$ = 0–0.9). The molar ratio of the flux was CsCl:KCl = 5:3. 0.8 g of weighed raw materials and 5.0 g of CsCl/KCl flux were mixed in a mortar and sealed in a quartz tube in vacuum. The mixed powder was heated at 800 °C for 10 h, cooled slowly to 600 °C at a rate of 1 °C/h, and then furnace-cooled to room temperature. The quartz tube was opened in air atmosphere, and the flux was dissolved in distilled water in the quartz tube. The products were filtered and washed with distilled water. The structures and compositions of the grown single crystals were evaluated by X-ray diffraction (XRD) analysis using CuKα radiation, scanning electron microscopy (SEM), and electron probe microanalysis (EPMA). The critical temperatures ($T_c$) of the grown single crystals were estimated from the magnetization–temperature ($M$–$T$) curve under zero-field cooling (ZFC) and field cooling (FC) using a superconducting quantum interface device (SQUID) with an applied magnetic field of 10 Oe parallel to the $c$-axis. The resistivity-temperature ($\rho$-$T$) and current-voltage ($I$-$V$) characteristics of the single crystals were measured by the standard four-probe method in the constant current ($J$) mode using a physical property measurement system (PPMS DynaCool; Quantum Design). The electrical terminals were made of silver paste. We measured the angular ($\theta$) dependence of resistivity ($\rho$) in the flux liquid state under various magnetic fields ($H$) and calculated superconducting anisotropy ($\gamma_s$) using the effective mass model.[25-27] For the $c$-axis transport measurement, we fabricated s-shaped junctions from a grown single crystal fixed on the SrTiO$_3$ single-crystal substrate by a three-dimensional (3D) focused ion beam (FIB) etching method using a Ga-ion beam.[23,24] The area of the junctions was about $5.4 \times 5.7$ μm$^2$ in the $ab$-plane. The thickness of the junctions was about 1.0 μm along the $c$-axis. Normal-state anisotropy ($\gamma_n$) was estimated from the $\rho$-$T$ characteristics of the as-grown ($J//ab$-plane) and fabricated ($J//c$-axis) samples. The sample-position (the angle between the $ab$-plane and the magnetic field) dependence of the flow voltage ($V_{ff}$) of the fabricated sample was measured under a magnetic field.

## 3. Results and discussion

Figure 1 shows a typical SEM image of a PrO$_{1-a}$F$_a$BiS$_2$ plate like single crystal, which is around 1–3 mm in size and 20–40 μm thick. The well-developed plane was determined to be the *ab*-plane by XRD measurement. Figure 2 shows the XRD pattern of a well-developed plane in a single crystal grown from a starting powder with a nominal composition of PrO$_{0.7}$F$_{0.3}$BiS$_2$. The peak positions are in agreement with those of the (00*l*) diffraction peaks of PrO$_{1-a}$F$_a$BiS$_2$ described in the literature.[6] The presence of only (00*l*) diffraction peaks of the PrO$_{1-a}$F$_a$BiS$_2$ structure indicates a well-developed *ab*-plane.

Figure 3 shows the *M–T* curve for single crystals grown from a starting powder with a nominal composition of PrO$_{0.7}$F$_{0.3}$BiS$_2$ under ZFC and FC. The Meissner effect was confirmed from the magnetization between 2 and 10 K. $T_c$ was defined in the separation temperature of FC and ZFC curves. The $T_c$ and superconducting volume fraction at 2.0 K were estimated to be 3.8 K and 75%, respectively.

Table I shows the F concentrations of the starting materials, *c*-axis lattice parameters, and $T_c$ and $\gamma_s$ values of the single crystals. The F concentrations in the starting materials and grown crystals are defined as *x* and *y*, respectively. Single crystals were obtained from the starting materials of $0.1 \leq x \leq 0.8$, but only a small amount of grown single crystals was obtained for $x = 0.1$. The chemical ratio of Pr:Bi:S in the single crystals was determined to be $1.06 \pm 0.02 : 1.02 \pm 0.03 : 2$ by EPMA, and the sulfur concentration was normalized to be 2. The composition of the grown crystals was almost stoichiometric, with a slightly Pr-rich composition. Cs, K, and Cl were not detected in the crystals by EPMA with a minimum sensitivity limit of 0.1 wt%. The value of $y$ in the grown crystals increased to up to 0.26 with increasing $x$ to up to 0.5, and it almost saturated at $x \geq 0.5$. The $c$-axis lattice parameters of the single crystals ($x > 0.2$) were close to those of polycrystalline samples.[28] The $c$-axis lattice parameters of the grown crystals decreased with increasing $x$, and they almost saturated at $x \geq 0.4$. These results suggest that the solubility limit of F in PrOBiS$_2$ crystals is between 0.4 and 0.5, as shown in Fig. 4. The values of $y$ and the $c$-axis lattice parameters are almost the same between $x = 0.5$ and 0.8. However, the $T_c$s show deviations of 4.1–4.6 K. We assume that the experimental error caused the deviations of $T_c$ values. Slight changes in the F concentration and/or local structure might affect the transition temperature.[17] The grown single crystals, except those at $x = 0.1$, showed superconducting transitions, and the F concentration in the PrO$_{1-a}$F$_a$BiS$_2$ grown crystals increased $T_c$ to up to around 4.5 K.

The $\theta$ dependence of $\rho$ was measured at different $H$ values in the flux liquid state to estimate the $\gamma_s$ values of $PrO_{1-a}F_aBiS_2$ single crystals in accordance with the experimental procedure in Refs. 25 and 26. The reduced field ($H_{red}$) was calculated using the following equation for an effective mass model:

$$H_{red} = H(\sin^2\theta + \gamma_s^{-2}\cos^2\theta)^{1/2}, \quad (1)$$

where $\theta$ is the angle between the $ab$-plane and the magnetic field,[27] and $H_{red}$ is calculated from $H$ and $\theta$. $\gamma_s$ was estimated from the best scaling of the $\rho$–$H_{red}$ relationship. Figure 5 shows the $\theta$ dependence of $\rho$ in the grown crystal of $y = 0.18$ under different magnetic fields ($H = 0.1$–9.0 T) in the flux liquid state at 3.0 K. The $\rho$–$\theta$ curve exhibits a two-fold symmetry. Figure 6 shows the $\rho$–$H_{red}$ scaling obtained from the $\rho$–$\theta$ curves in Fig. 5 using Eq. (1). The scaling was performed by taking $\gamma_s = 58$, as shown in Fig. 6. Table I shows the superconducting anisotropies of $PrO_{1-a}F_aBiS_2$ single crystals with various F concentrations. The $\gamma_s$ values of the single crystals were calculated to be 20–58, and the high values of 53–58 were found in the single crystals of $x = 0.3$. This value is higher than that of a $YBa_2Cu_3O_a$ (Y-123) high-$T_c$ cuprate superconductor ($\gamma_s = 45.8$), which shows intrinsic Josephson junctions.[29] Thus, the $PrO_{1-a}F_aBiS_2$ single crystals of $x = 0.3$ were employed for the fabrication of s-shaped intrinsic Josephson junctions, as described below.

Figure 7 shows a scanning ion microscopy (SIM) image of the s-shaped junctions fabricated on the PrO$_{1-a}$F$_a$BiS$_2$ single crystals of $x$ = 0.3 along the $c$-axis. The cross-sectional area and thickness of the junctions are 5.4 × 5.7 μm$^2$ along the $ab$-plane and 1.0 μm, respectively. The direction of current flow is shown in Fig. 7, which is along the $c$-axis in the junction. This junction was used for characterizing the transport properties along the $c$-axis. Figure 8 shows the $\rho$-$T$ characteristics along the (a) $ab$-plane ($\rho_{ab}$) and (b) $c$-axis ($\rho_c$) of a Pr$_{1.05}$O$_{0.82}$F$_{0.18}$Bi$_{1.03}$S$_{2.00}$ ($y$ = 0.18) single crystal, as measured by a standard four-probe method. The zero-resistivity temperature ($T_c^{zero}$) along the $ab$-plane was determined to be 2.95 K, whereas that along the $c$-axis decreased slightly to 2.70 K. Ga ions were likely implanted in the fabricated sample during the FIB etching, which possibly degraded the sample. This decrease in $T_c^{zero}$ may be attributed to the damage caused by FIB etching. The normal-state anisotropies $\gamma_n$ = ($\rho_c/\rho_{ab}$)$^{1/2}$ for Pr$_{1.05}$O$_{0.82}$F$_{0.18}$Bi$_{1.03}$S$_{2.00}$ single crystals at 5 and 250 K were found to be 39.3 and 21.1, respectively. These results suggest that this layered compound shows a highly anisotropic transport even in the normal state. These $\gamma_n$ values were lower than the $\gamma_s$ values. The $\rho$–$T$ relationship along the $ab$-plane ($J$//$ab$-plane) exhibits an anomaly at approximately 140 K and small kinks at ~70 K. The former anomaly is similar to those observed in NdO$_{1-a}$F$_a$BiS$_2$ and LaO$_{1-a}$F$_a$BiSe$_2$ single crystals,[13,30] in which temperature may be related to F concentration.[30] Thus, those anomalies would be intrinsic properties of $BiS_2$- and $BiSe_2$-based superconductors.[30] The small kinks at ~70 K along the *ab*-plane (*J*//*ab*-plane) and *c*-axis (*J*//*c*-axis) are reproducible; however, in origin is unclear.

Figure 9 shows the *I–V* characteristics along the *c*-axis of the $Pr_{1.05}O_{0.82}F_{0.18}Bi_{1.03}S_{2.00}$ single crystal (shown in Fig. 7) at 2.0 K when $T/T_c$ was around 0.74. The critical current density ($J_c$) was approximately $1.33 \times 10^3$ A/cm$^2$ in the self-field. The *I–V* curve shows a hysteresis. However, the multi branch structure of the *I–V* characteristics that corresponds to intrinsic Josephson junctions[23,24] was not observed in the s-shaped $Pr_{1.05}O_{0.82}F_{0.18}Bi_{1.03}S_{2.00}$ single-crystal junctions.

We measured the sample-position (the angle between the *ab*-plane and the magnetic field) dependence of the $V_{ff}$ characteristics under a magnetic field (*H*) for s-shaped $Pr_{1.05}O_{0.82}F_{0.18}Bi_{1.03}S_{2.00}$ single-crystal junctions (Fig. 10). We defined flow resistance as $V_{ff}$ divided by dc bias current (*I*) to conduct the measurements. Figure 10 shows the sample-position dependence of the $V_{ff}$ characteristics at various *I*s along the *c*-axis of the s-shaped junctions in a 1.0 T magnetic field at 2.0 K. Figure 10(b) shows that $V_{ff}$ reaches a local maximum under a magnetic field parallel to the *ab*-plane. The local maximum $V_{ff}$ increases with increasing applied current. An increase in the local maximum $V_{ff}$ is also observed in the $I$–$V$ characteristics under a magnetic field parallel to the $ab$-plane, as shown in Fig. 10(c). This phenomenon can be explained by the "lock-in" state,[31,32] which may originate from the Josephson vortex flow. We expect that pancake vortices appear in $Pr_{1.05}O_{0.82}F_{0.18}Bi_{1.03}S_{2.00}$ single crystals owing to the high superconducting anisotropy. $V_{ff}$ markedly decreases under magnetic fields approximately parallel to the $ab$-plane, suggesting the dissipation of pancake vortices. Subsequently, Josephson vortices appear under magnetic fields parallel to the $ab$-plane, indicating the "lock-in" state. The lock-in state is expected to be free from pancake vortices, which cross the superconducting $BiS_2$ layers. We observed periodic oscillations of Josephson-vortex flow resistance.[29,33,34] The starting field of the periodic oscillations ($H_s$) was calculated from the following equation:[29,35]

$$H_s = \Phi_0/2\pi\lambda_j s, \qquad (2)$$

where $\lambda_j = \gamma_s s$ expresses the size of the nonlinear Josephson-vortex core along the $c$-axis (frequently called the "Josephson penetration depth $\lambda_j$"), $s$ is the length of a unit cell (= $c$-axis lattice constant of $Pr_{1.05}O_{0.82}F_{0.18}Bi_{1.03}S_{2.00}$), and $\Phi_0$ is the flux quantum (= 2.07 × $10^{-15}$ Wb). $H_s$ was calculated from Eq. (2) and was estimated to be 3.12 T for the s-shaped $Pr_{1.05}O_{0.82}F_{0.18}Bi_{1.03}S_{2.00}$ single-crystal junctions. However, no oscillation was observed until 7.0 T at 2.0 K. The absence of oscillation may be attributed to the measurement temperature; $T/T_c$ is around 0.74 at 2.0 K. Although our results demonstrate that $Pr_{1.05}O_{0.82}F_{0.18}Bi_{1.03}S_{2.00}$ single crystals are promising candidates for use as intrinsic Josephson junctions, further investigations at lower temperatures are required.

## 4. Conclusions

We grew $PrO_{1-a}F_aBiS_2$ single crystals with well-developed *ab*-planes 1–3 mm in size using CsCl/KCl flux. Increasing the F concentrations of the starting materials enhanced the substitution of F in the grown crystals. The superconducting $T_c$ and $\gamma_s$ values of $PrO_{1-y}F_yBiS_2$ ($y$ = 0.05–0.28) single crystals were estimated to be 2.4–4.6 K and 20–58, respectively. The normal-state anisotropies ($\gamma_n$) of $Pr_{1.05}O_{0.82}F_{0.18}Bi_{1.03}S_{2.00}$ ($x$ = 0.3, $y$ = 0.18) single crystals at 5 and 250 K were found to be 39.3 and 21.1, respectively. The s-shaped $Pr_{1.05}O_{0.82}F_{0.18}Bi_{1.03}S_{2.00}$ single-crystal junctions under a magnetic field parallel to the *ab*-plane exhibited a lock-in state at 2.0 K.

**Figure captions**

Fig. 1. Typical SEM image of PrO$_{1-a}$F$_a$BiS$_2$ single crystal. The well-developed plane is the *ab*-plane.

Fig. 2. XRD pattern of well-developed plane of PrO$_{1-a}$F$_a$BiS$_2$ single crystal grown from starting powder with nominal composition of PrO$_{0.7}$F$_{0.3}$BiS$_2$.

Fig. 3. *M–T* curve for single crystals grown from starting powder with nominal composition of PrO$_{0.7}$F$_{0.3}$BiS$_2$ under ZFC and FC.

Fig. 4. Dependence of starting materials' nominal F composition *x* on (a) analytical F composition *y* and (b) *c*-axis lattice parameter.

Fig. 5. Angular $\theta$ dependence of resistivity $\rho$ in flux liquid state at various magnetic fields for Pr$_{1.05}$O$_{0.82}$F$_{0.18}$Bi$_{1.03}$S$_{2.00}$ single crystal.

Fig. 6. Same data as in Fig. 5; scaling of angular $\theta$ dependence of resistivity $\rho$ at reduced magnetic field $H_{\text{red}} = H(\sin^2\theta + \gamma_s^{-2}\cos^2\theta)^{1/2}$.

Fig. 7. SIM image of s-shaped junctions. The cross-sectional area and thickness of the junctions are about 30.78 μm$^2$ and 1.0 μm, respectively.

Fig. 8. $\rho$–*T* characteristics of Pr$_{1.05}$O$_{0.82}$F$_{0.18}$Bi$_{1.03}$S$_{2.00}$ single crystal along the (a) *ab*-plane ($\rho_{ab}$) and (b) *c*-axis ($\rho_c$). Sample for *c*-axis transport measurement is shown in Fig. 7.

Fig. 9. I–V characteristics of s-shaped $Pr_{1.05}O_{0.82}F_{0.18}Bi_{1.03}S_{2.00}$ single-crystal junctions at 2.0 K and self-field. $J_c$ is about $1.33\times10^3$ A/cm$^2$.

Fig. 10. (a) Sample-position (angle between *ab*-plane and magnetic field) dependence of flow voltage $V_{ff}$ at various currents (*I*) for s-shaped $Pr_{1.05}O_{0.82}F_{0.18}Bi_{1.03}S_{2.00}$ single-crystal junctions under 1.0 T magnetic fields at 2.0 K. (b) Enlargement of sample position approximately parallel to *ab*-plane. (c) I–V characteristics of s-shaped junctions at 2.0 K and 1.0 T magnetic field parallel to *ab*-plane.

Table I. Dependence of nominal F concentration in the starting materials ($x$) on the analytical F concentration ($y$), $c$-axis lattice parameter ($c$), superconducting transition temperature ($T_c$), and superconducting anisotropy ($\gamma_s$) in the grown single crystals. The ratios of F, $x$, and $y$ are determined by F/(O+F).

| $x$ | $y$ | $c$ (Å) | $T_c$ (K) | $\gamma_s$ |
|-----|-----|---------|-----------|------------|
| 0   | _a) | _a)     | _a)       | _a)        |
| 0.1 | 0.05 | 13.67  | _b)       | _b)        |
| 0.2 | 0.13 | 13.55  | 2.4       | 20         |
| 0.3 | 0.18 | 13.49  | 3.8       | 53-58      |
| 0.4 | 0.23 | 13.41  | 4.0       | 32-46      |
| 0.5 | 0.26 | 13.39  | 4.1       | 42-56      |
| 0.6 | 0.26 | 13.38  | 4.3       | 47         |
| 0.7 | 0.26 | 13.37  | 4.4       | 38-39      |
| 0.8 | 0.28 | 13.37  | 4.6       |            |
| 0.9 | _a) | _a)     | _a)       | _a)        |

a): No PrO$_{1-a}$F$_a$BiS$_2$ single crystals were obtained. b): Unmeasurable in our system and small crystals.

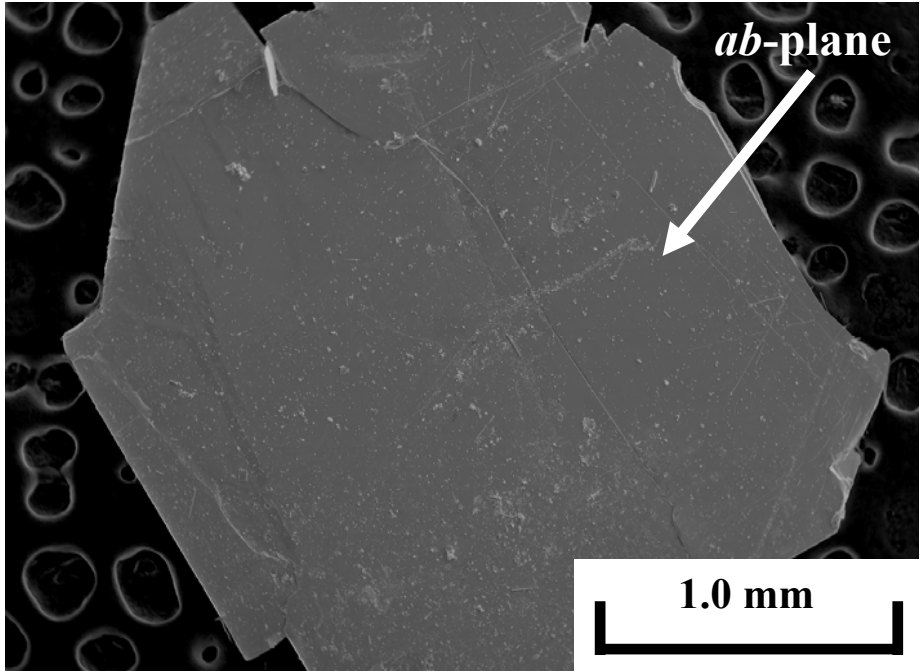

**Fig. 1**

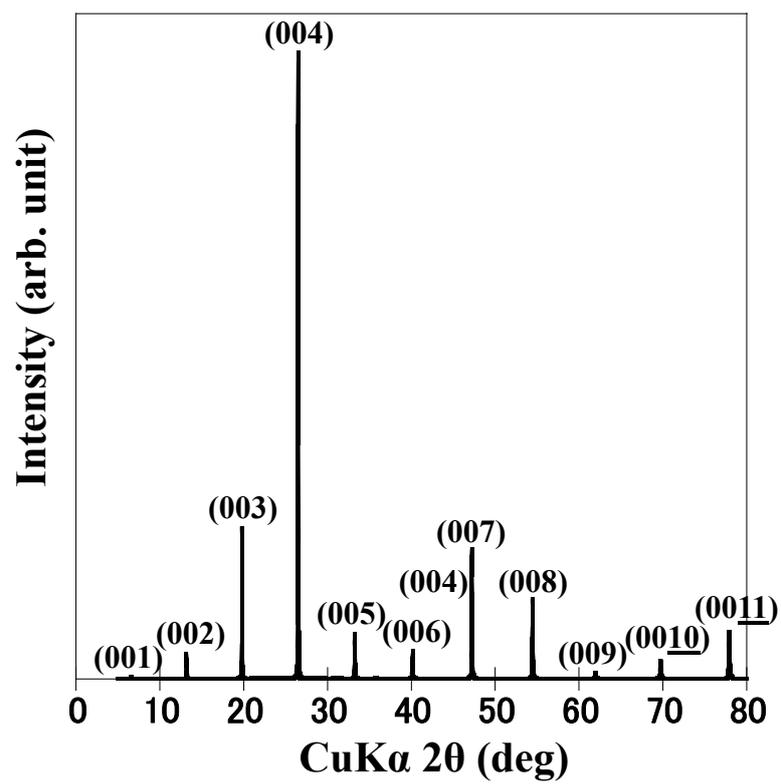

Fig. 2

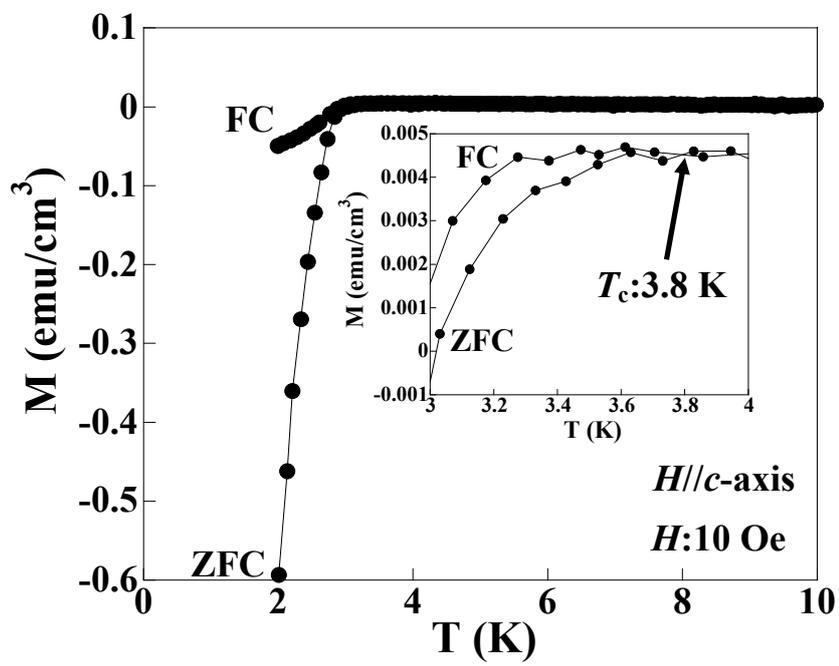

Fig. 3

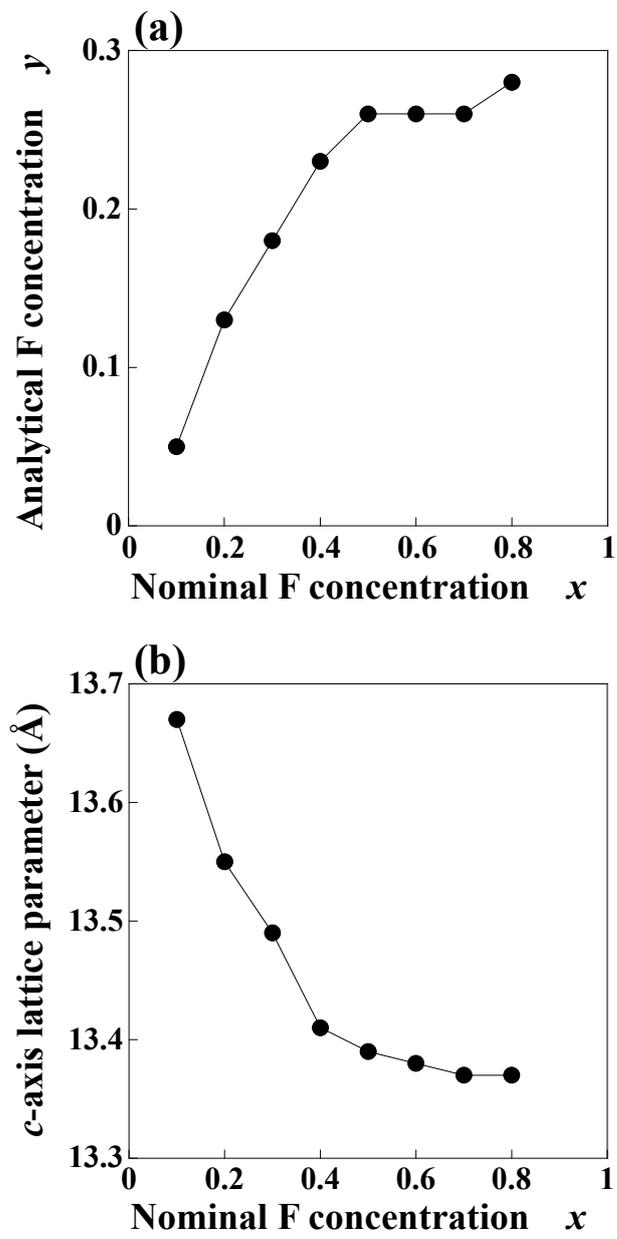

**Fig. 4**

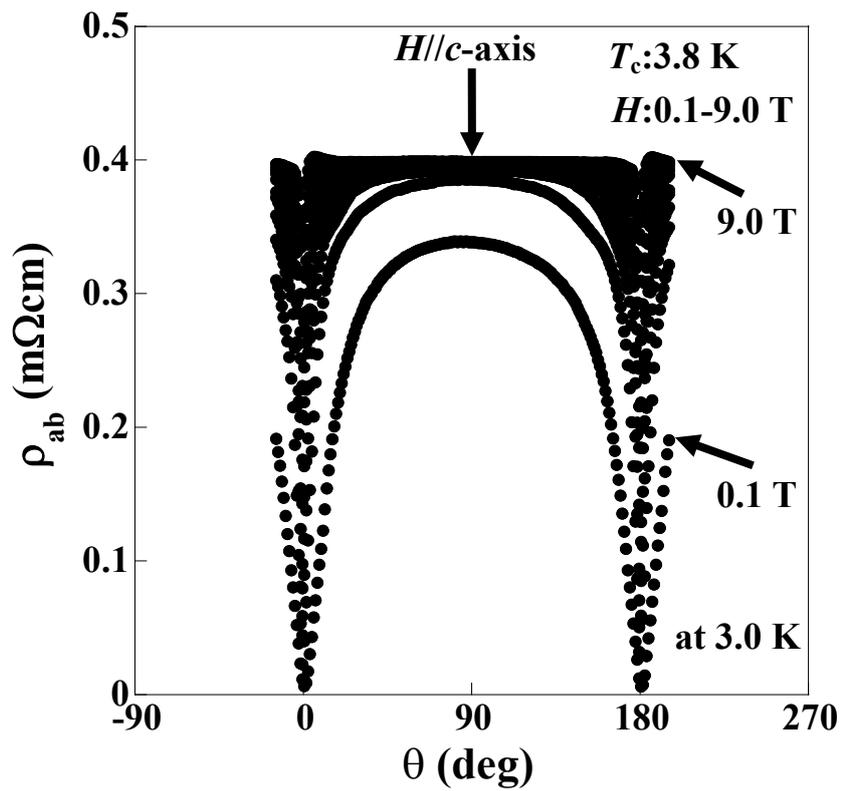

**Fig. 5**

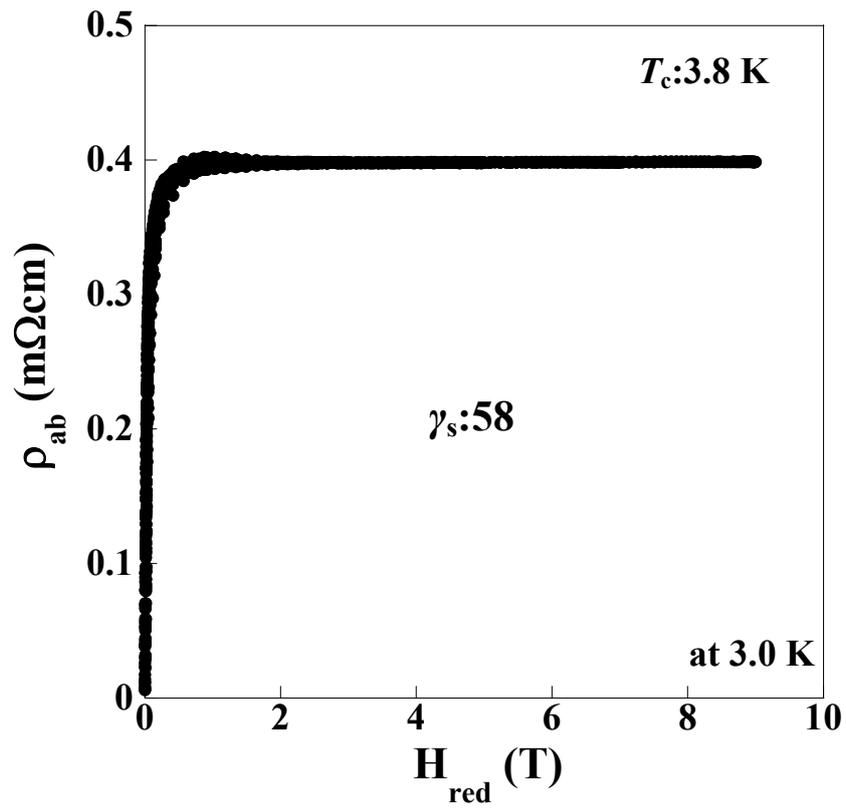

Fig. 6

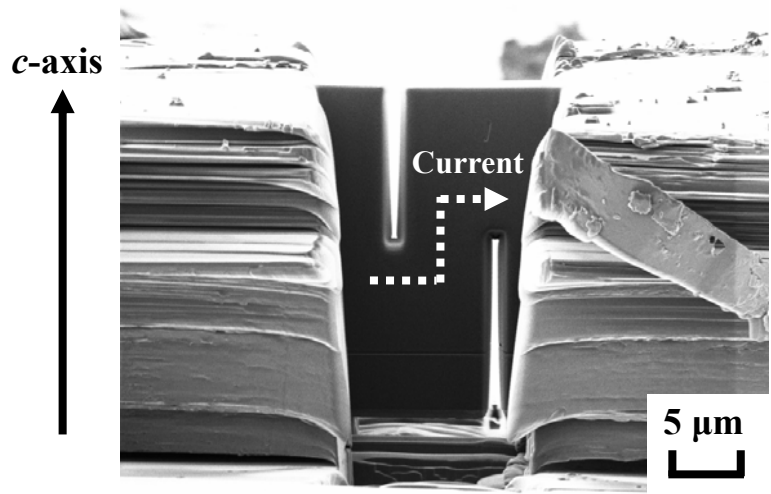

**Fig. 7**

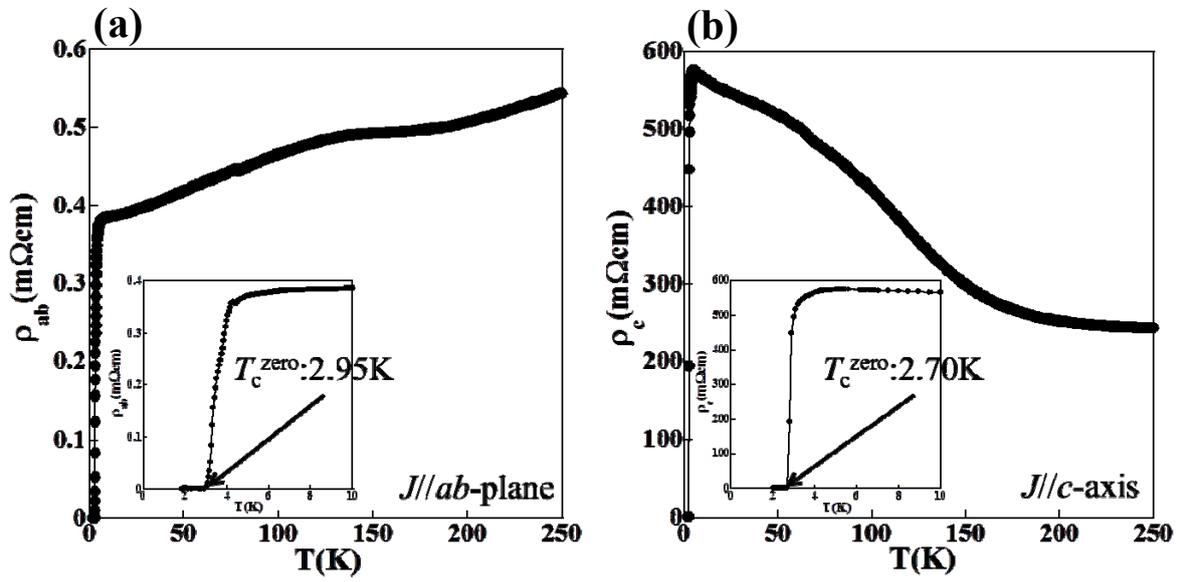

**Fig. 8**

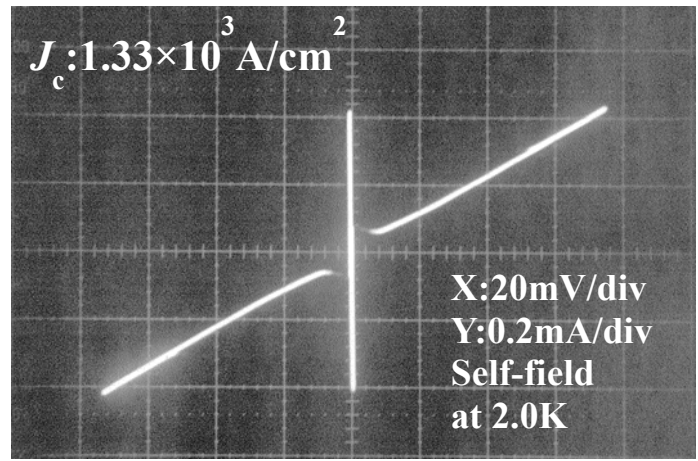

**Fig. 9**

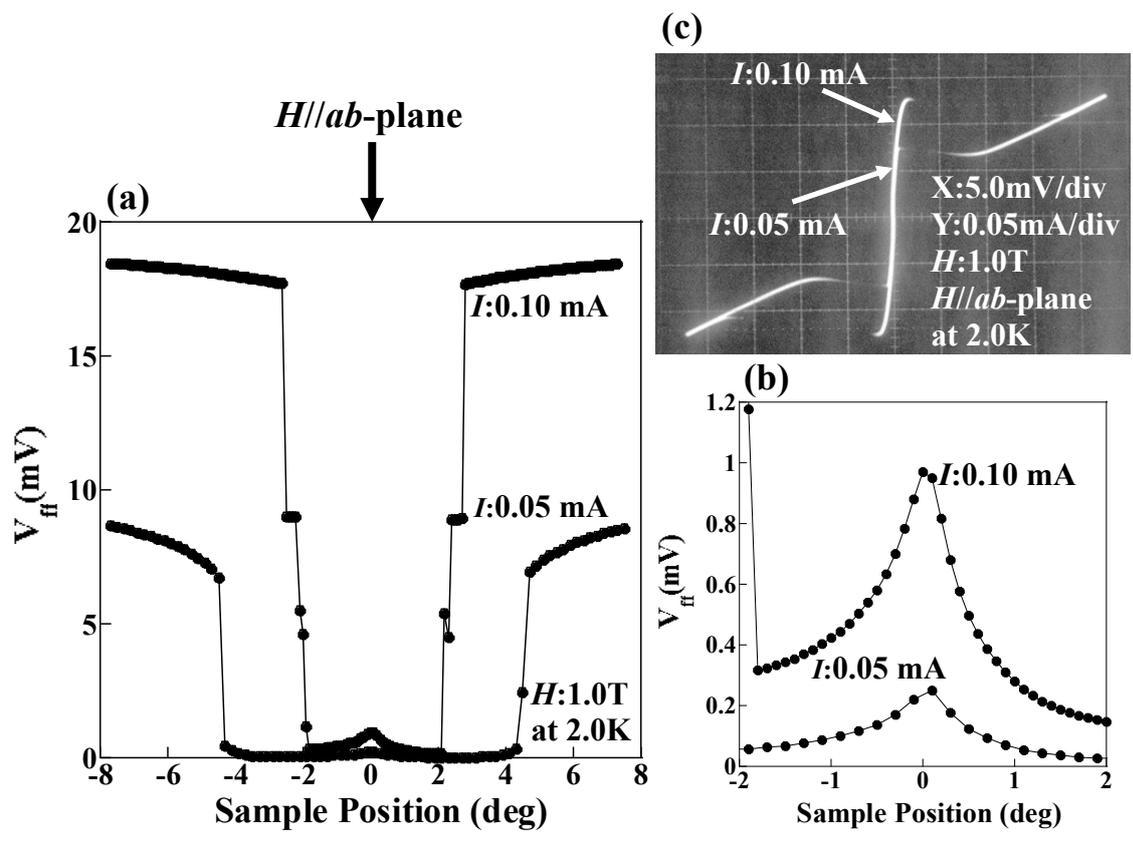

**Fig. 10**